\documentclass[preprint, nofootinbib]{revtex4}
\usepackage{graphicx}
\usepackage{amsmath,amsthm,amssymb}

\begin{document}
%


\title{Huygens' principle and anomalously small radiation tails}

\author{Piotr Bizo\'n}
\affiliation{M. Smoluchowski Institute of Physics, Jagiellonian
University, Krak\'ow, Poland}

\date{\today}
\begin{abstract}
This is a short account of recent joint work with T. Chmaj and A. Rostworowski on asymptotic
behavior of linear and nonlinear waves, as presented at the conference devoted to Myron Mathisson
held at the Banach Center, Warsaw 2007.
\end{abstract}

\maketitle

\section{Introduction}
The work presented in this talk is part of a long-time project aimed at the detailed quantitative
description of the process of relaxation to equilibrium for nonlinear wave equations defined on
spatially unbounded manifolds. By equilibrium we mean here a stable stationary solution, like a
soliton, a black hole, or just a flat space. The convergence to these solutions occurs through a
mechanism of radiating an excess energy to infinity. For a large class of physically interesting
systems the late stages of this process are universal: for intermediate times the convergence has
the form of exponentially damped oscillations (called quasinormal modes) and asymptotically it
has the form of polynomial decay (called a tail). This very last stage of the relaxation process,
the tail, is the subject of my talk.

  The presentation of this talk at the conference devoted to Myron Mathisson is justified by the
  fact our results shed new light on Huygens' principle, one of the main subjects of Mathisson's
  mathematical interests. Recall that a
wave equation is said to satisfy Huygens' principle if: (i) the solution at a point $P$ depends
only on the initial data at the intersection of the past light cone of $P$ with the Cauchy
hypersurface or,  equivalently, (ii)  the solution vanishes at all points which cannot be reached
from the initial data by a null geodesic (i.e., there is no tail). A prototype equation
satisfying Huygens' principle is the ordinary wave equation in $d+1$ dimensional Minkowski
spacetime for odd $d\geq 3$. Actually, according to  Hadamard's conjecture  \cite{ch} this is the
only (modulo trivial transformations) huygensian linear second-order hyperbolic equation of the
form
\begin{equation}
g^{\mu\nu}(x) \nabla_{\mu} \nabla_{\nu} \phi + A^{\mu}(x)\nabla_{\mu} \phi + B(x) \phi=0\,.
\end{equation}
 Mathisson
proved this conjecture in the case of four dimensional Minkowski spacetime \cite{m}.
Counterexamples to Hadamard's conjecture, which have been found later (see \cite{g} and Roy
McLenaghan's  talk at this conference), do not change the fact that Huygens' property is a very
rare and unstable phenomenon. Thus, it is natural to ask if there are perturbations of the free
wave equation which preserve Huygens' property approximately, in the sense that the tail which is
left behind the wave front is very small. The existence of such special perturbations in higher
even dimensions is a byproduct of our studies of tails.
\section{Model and assumptions}
We consider equations of the form
\begin{equation}\label{main}
\Box \phi+V(x)\phi +N(\phi,\nabla\phi,x)=0\,,\qquad
 \Box=\partial_t^2-\Delta\,,\quad (t,x)\in R^{d+1}\,,
 \end{equation}
for spherically symmetric smooth initial data with compact support. Since we want the free part
to satisfy Huygens' property, we restrict ourselves  to odd spatial dimension $d\geq 3$. Apart
from obvious mathematical motivations, there are at least two physical reasons for studying
higher dimensions $d>3$. First, for linear wave equations higher dimensions are equivalent to
higher spherical harmonics. This follows from the identity
\begin{equation}
\left(\partial_t^2-\partial_r^2-\dfrac{d-1}{r} \partial_r\right)\phi
=\dfrac{1}{r^l}\left(\partial_t^2-\partial_r^2-\dfrac{2}{r}
\partial_r+\dfrac{l(l+1)}{r^2}\right)(r^l
\phi)\,,\quad \qquad l=(d-3)/2\,,
\end{equation}
which relates the $l=0$ radial wave operator in $d$ space dimensions with the radial wave
operator in three space dimensions for the $l$th spherical harmonic with $l=(d-3)/2$.

Second, some geometric wave equations in $3+1$ dimensions are equivalent to scalar wave equations
in $d+1$ dimensions for $d>3$. For example, equivariant wave maps from $R^{3+1}$ into $S^3$
satisfy the following equation
\begin{equation}
\left(\partial_t^2-\partial_r^2-\dfrac{2}{r}\partial_r\right)\psi+\dfrac{\sin(2\psi)}{r^2} =0\,,
\end{equation}
which, after substitution $\psi=r\phi$, becomes the nonlinear scalar wave equation in $5+1$
dimensions
\begin{equation}
\left(\partial_t^2-\partial_r^2-\dfrac{4}{r}\partial_r\right)\phi+\dfrac{4}{3} \phi^3
+\mbox{higher order terms} =0\,.
\end{equation}

\section{Tools}
In this section we recall two elementary tools from the theory of linear wave equations. The
first tool is a formula  for the spherically symmetric solution of the free wave equation $\Box
\phi =0$ in $d+1$ dimensions (hereafter we shall use $l=\dfrac{d-3}{2}$ instead of $d$):
\begin{equation}\label{tool1}
    \phi(t,r)= \frac{1}{r^{2l+1}}\,\sum_{k=0}^{l}  \frac {2^{k-l}(2l-k)!} {k!(l-k)!}
   \, r^k
\left(a^{(k)}(t-r)-(-1)^k a^{(k)}(t+r)\right)\,.
\end{equation}
This solution, which is a superposition of ingoing and outgoing waves, is parameterized by a
single function $a(r)$ uniquely determined by initial data (the superscript in round brackets
denotes the $k$-th derivative).

The second tool is the Duhamel formula for the solution of the inhomogeneous free wave equation
$\Box \phi=F(t,r)$ with zero data
\begin{equation}\label{duhamel}
\phi(t,r)= \frac{1}{2 r^{l+1}}
    \int\limits_{0}^{t} d\tau \int\limits_{|t-r-\tau|}^{t+r-\tau}
    \rho^{l+1} P_l(\mu)\ F(\tau,\rho) d\rho\,,\quad  \qquad \mu=\frac{r^2+\rho^2-(t-\tau)^2}{2 r
    \rho}\,,
\end{equation}
    where $P_l(\mu)$ is the Legendre polynomial of
    degree $l$
This expression  can be easily obtained from the standard Green's function formula by integrating
out the angular variables \cite{ls}.
    In terms of null coordinates $u=\tau-\rho$ and $v=\tau+\rho$ the Duhamel formula (\ref{duhamel})
    takes a more convenient form
\begin{equation}\label{tool2nul}
\phi(t,r)= \frac{1}{2^{l+3}  r^{l+1}}
    \int\limits_{|t-r|}^{t+r} dv \int\limits_{-v}^{t-r}
    F(u,v) (v-u)^{l+1} P_l(\mu) du\,, \qquad \mu=\frac{r^2+(v-t)(t-u)}{r(v-u)}\,.
    \end{equation}
The formulae (\ref{tool1}) and (\ref{tool2nul}) will be used repeatedly below.

\section{Linear tails}
For the clarity of presentation we first consider the linear equation with a potential
\begin{equation}\label{eq_linear}
 \Box \phi
+\lambda V \phi =0\,, \qquad (\phi(0,r),\partial_t\phi(0,r))=(f(r),g(r))\,.
\end{equation}
The prefactor $\lambda$, introduced for convenience, will be assumed small and used as a
perturbation parameter. In order to determine the long-time behavior of $\phi(t,r)$ we define the
perturbation series
\begin{equation}\label{ser_lin}
\phi=\phi_0+\lambda \phi_1+\lambda^2\phi_2 + ...\,,
\end{equation}
where $\phi_0$ satisfies initial data (\ref{eq_linear}) and all higher $\phi_n$ have zero data.
Substituting this series  into equation (\ref{eq_linear}) we get the iterative scheme
\begin{equation}
  \Box \phi_0 = 0,\quad
  \Box  \phi_1 = -V \phi_0,\quad
  \Box \phi_2 = -V \phi_1,\quad \mbox{etc.}
\end{equation}
which can be solved recursively using the formulae (\ref{tool1}) and (\ref{tool2nul}). Assuming
that $V(r)\sim r^{-\alpha}$ ($\alpha>2$) for $r \rightarrow \infty$, we showed in \cite{bcr} that
 the leading order asymptotic behavior at timelike infinity  (fixed $r$ and $t\rightarrow
 \infty$) is given by
\begin{equation}\label{tail_lin}
 \phi_1(t,r) = \frac {C(l,\alpha)} {t^{\alpha+2l}}
  \left[ A + \mathcal{O} \left( \frac{1}{t}
\right) \right]\,,
\end{equation}
where\footnote{Here we use the notation:
    \begin{eqnarray}
x^{\underline{0}} := 1, &\qquad& x^{\underline{k}} := x \cdot (x-1) \cdot \dots \cdot (x-(k-1)),
\quad k>0\,,\nonumber
\\
x^{\overline{0}} := 1, &\qquad& x^{\overline{k}} := x \cdot (x+1) \cdot \dots \cdot (x+(k-1)),
\quad k>0\,.\nonumber
\end{eqnarray}}
\begin{equation}\label{C}
        C(l,\alpha) = - \frac {2^{\alpha+2l-1}}{(2l+1)!!}
        \left( \frac {\alpha-3} {2} \right)^{\underline{l}}
        \left( \frac {\alpha} {2} \right)^{\overline{l}}\quad
        \mbox{and} \quad
        A=\int \limits_{-\infty}^{+\infty} a(u)\, du\,.
        \end{equation}
        The constant $A$ is the only trace of initial data.
 The expression (\ref{tail_lin}) was  first derived by Ching et al. \cite{ching} who used Fourier
 transform methods.

We claim that the first iterate provides a good approximation of the entire tail if $\lambda$ is
sufficiently small, that is
\begin{equation}\label{asym}
\phi(t,r)-\lambda \phi_1(t,r) \sim \mathcal{O}(\lambda^2) t^{-(\alpha+2l)}\,.
\end{equation}
This basically follows from the fact that all higher-order iterates
 $\phi_n(t,r)$
decay in the same manner (or faster) as $\phi_1(t,r)$. Of course, the main issue is whether the
perturbation series is convergent; for $l=0$ (i.e., in three space dimensions) this was proved in
\cite{sbcr}) but for higher $l$ the problem is open. Note however that for practical purposes it
is sufficient that the series is asymptotic to the solution.

The numerical verification of (\ref{asym}) shows perfect agreement with analytic predictions
\cite{bcr}. We remark that numerical simulations of tails are not quite trivial even in the
radial case because discretization errors generate  artificial tails which might mask the true
behavior. To eliminate such artifacts one has to use high-order finite difference schemes. In
addition, quadruple precision is needed to suppress the accumulation of round-off errors during
long-time simulations. For these reasons the simulations of tails are computationally expensive.
\section{Nonlinear  tails} In this section we consider equation (\ref{main}) without a potential. For
simplicity, we take a pure power nonlinearity with an integer exponent $p\geq 3$ (the
generalization to other nonlinearities is straightforward) \begin{equation}\label{eq_non}
 \Box \phi -\phi^p =0\,, \qquad
(\phi(0,r),\partial_t\phi(0,r))=(\varepsilon f(r),\varepsilon g(r))\,.
\end{equation}
 This time the amplitude of initial data $\varepsilon$ plays the role of a small parameter in
the perturbation series:
\begin{equation}\label{ser_non}
\phi=\varepsilon \phi_0+\varepsilon^2 \phi_1+\varepsilon^3\phi_2 + ...
\end{equation}
Substituting (\ref{ser_non}) into equation (\ref{eq_non}) we get the iteration scheme
\begin{equation}
  \Box \phi_0 = 0,\quad
  \Box  \phi_p = \phi_0^p,\quad \mbox{etc.}
\end{equation}
As above, we get $\phi_0$ using the formula (\ref{tool1}) and then evaluate $\phi_p$ using the
Duhamel formula (\ref{tool2nul}). In the limit of timelike infinity we obtain \cite{bcr2}
\begin{equation}\label{tail_non}
\phi_p(t,r) = \frac {\tilde C(l,p)} {t^{(l+1)p-1}} \left[ \tilde A + \mathcal{O} \left(
\frac{1}{t} \right) \right]\,,
\end{equation}
where
\begin{equation}\label{C2}
        \tilde C(l,p) = (-1)^l\, \frac {2^{(l+1)(p+1)-1}}{(2l+1)!!} [(l+1)(p-1)-2]^{\underline{l}}
        \quad
        \mbox{and} \quad
        \tilde A=\int \limits_{-\infty}^{+\infty} [a^{(l)}(u)]^p\, du\,.
    \end{equation}
The remarks given above in the linear case apply verbatim to the nonlinear case as well; in
particular the perturbation series (\ref{ser_non}) is known to converge for $l=0$ \cite {sbcr}
and is at least asymptotic to the full solution for $l>0$.

\section{Competition between linear and nonlinear tails} The most interesting situation occurs
when both the potential and the nonlinearity are present in equation (\ref{main}). Then, each of
these terms produces its own tail:
\begin{equation}\label{comp}
\mbox{linear tail} \sim t^{-(\alpha+2l)}\quad vs.\quad \mbox{nonlinear tail} \sim t^{1-(l+1)p}\,.
\end{equation}
Clearly, the tail with slower decay rate is dominant asymptotically, that is
\begin{equation}\label{compet}
\phi(t,r) \sim t^{-\gamma}\,, \qquad \gamma=\min\{\alpha+2l,(l+1)p-1\}\,.
\end{equation}
For $l=0$ this result (without a coefficient) was first proved by Strauss and Tsutaya \cite{st}.

To give an example of the competition of linear and nonlinear tails, let us consider the Skyrme
model. Under a spherical symmetry reduction (corotational ansatz), this model reduces to the
single nonlinear wave equation for the function $F(t,r)$
\begin{equation}\label{skyrme}
\partial_t(w \partial_t F) - \partial_r(w\partial_r F)
 + \sin(2F) + \sin(2F) \left(\frac{\sin^2{F}}{r^2}\!+\! (\partial_r F)^2-
 (\partial_t F)^2\right)=0\,,
 \end{equation}
 where $w=r^2+2
\sin^2{F}$. Regular finite energy solutions of (\ref{skyrme}) must satisfy the boundary
conditions $F(t,0)=0$ and $F(t,\infty)=m\pi$, where an integer $m$ has the interpretation of the
topological degree of the solution. For $m=1$, equation (\ref{skyrme}) has a regular static
solution
 $S(r)$ called the skyrmion. This solution is linearly stable and plays the role of a global attractor, that is, every solution
 starting from smooth finite energy initial data of degree one remains globally regular for all
 times and  asymptotically converges to $S(r)$.
 The perturbation $\phi(t,r)=\sqrt{w} (F(t,r)-S(r))/r^2$ satisfies the equation
 \begin{equation}
\left(\partial_t^2-\partial_r^2-\dfrac{4}{r}\partial_r\right)\phi+V(r)\phi+ \dfrac{4}{3} \phi^3
+\mbox{higher order terms} =0\,,
\end{equation}
where the potential $V(r)$ has no bound states and falls off as $r^{-6}$ for $r\rightarrow
\infty$. For this equation we have $d=5\, (l=1), \,\,\alpha=6$ and  $p=3$, hence from
(\ref{comp})
\begin{equation}
\mbox{linear tail} \sim t^{-8}\qquad \mbox{and} \qquad \mbox{nonlinear tail} \sim t^{-5}\,.
\end{equation}
Thus, the nonlinear tail is dominant \cite{skyrme}. This example shows that one has to be
cautious in drawing conclusions about the asymptotic behavior of solutions of nonlinear wave
equations on the basis of linear perturbation analysis -- even for small amplitude solutions the
nonlinear effects can be dominant.
\section{Anomalous tails}
An advantage of our approach, in contrast to decay estimates in the form of inequalities, is that
we control the coefficient of the leading order term of the tail. This allows us to identify
those exceptional cases in which this coefficient vanishes and the decay is faster. We shall
refer to such tails as anomalous.

Let us first consider the linear case with the pure inverse power potential near infinity, that
is $V(r)=\lambda r^{-\alpha}$ for $r>R$. Then, it follows from (\ref{C}) that $C(l,\alpha)
\propto \left( \dfrac {\alpha-3} {2} \right)^{\underline{l}}=0$ if $\alpha$ is an odd integer
$\leq2l+1$, hence there is no tail in the first order. This means that the system
\begin{equation}
\Box \phi_0=0,\qquad  \Box \phi_1=-V \phi_0
\end{equation}
is huygensian.
 In order to find the tail in this exceptional case we need to solve  the second iteration
 equation $\Box \phi_2=-V \phi_1$ via the Duhamel formula. After a long calculation (which
 requires
  the asymptotic expansion of $\phi_1$ at null infinity) we get
(see \cite{bcr} for the details)
\begin{equation}
 \phi(t,r) \approx \lambda^2 \phi_2(t,r) =
  \lambda^2 \frac {D(l,\alpha)} {t^{2(\alpha+l-1)}} \left[ A + \mathcal{O} \left( \frac{1}{t}
\right) \right]\,,
\end{equation}
where the coefficient $D(l,\alpha)$ is given by a complicated but explicit expression (see
Eq.(23) in \cite{bcr}).

Next, consider the pure power nonlinearity $\Box \phi=\phi^p$. It follows from (\ref{C2}) that
$\tilde C(l,p) \propto [(l+1)(p-1)-2]^{\underline{l}}=0$ if $p=2$ and $l\geq 1$. Thus, in higher
even dimensions the first order tail vanishes for the quadratic nonlinearity. This implies that
the system \begin{equation} \Box \phi_0=0,\qquad \Box \phi_1=\phi_0^2\,
\end{equation}
 is huygensian. As before, the leading order behavior of the tail can be obtained by
solving the second order equation $\Box \phi_2=2\phi_0 \phi_1$ via the Duhamel formula. The
result (see \cite{bcr2} for the details) is
\begin{equation}\label{anom}
\phi(t,r)\approx \varepsilon^3 \phi_2(t,r)\sim \varepsilon^3 \frac{c(l)}{t^{3l+1}}\,,\qquad
c=(-1)^l
    \frac{2^{3l}}{2l(2l+1)} \int_{-\infty}^{\infty} a^{(l-1)}(\eta)[a^{(l)}(\eta)]^2 d\eta\,.
\end{equation}
Note that quadratic nonlinearities occur frequently in nonlinear perturbation theory so anomalous
tails are in fact quite common. As an example, consider the Yang-Mills field in four dimensions
with the $SO(3)$ gauge group, so that the potential $A_{\alpha}(x)$ is the skew-symmetric
$3\times 3$ matrix $A^{ij}_{\alpha}(x)$. For the spherically-symmetric ansatz
\begin{equation}\label{ansatz}
A^{ij}_{\mu}(x) = \left(\delta^j_{\mu}x^i-\delta^i_{\mu}x^j\right) \phi(t,r)\,,
\end{equation}
the Yang-Mills equation  $\partial_{\alpha}F^{\alpha\beta}+ [A_{\alpha}, F^{\alpha\beta}] = 0$,
where $F_{\alpha\beta}=\partial_{\alpha}A_{\beta}-\partial_{\beta}A_{\alpha}+
[A_{\alpha},A_{\beta}]$, reduces to the scalar semilinear wave equation in $5+1$ dimensions
\begin{equation}\label{ym}
(\partial_t^2 -\partial_r^2  -\frac{4}{r} \partial_r) \phi + 3\phi^2 + r^2 \phi^3=0\,.
\end{equation}
The quadratic term in (\ref{ym}) produces an anomalous tail (\ref{anom}) which is of the same
order, $\mathcal{O}(\varepsilon^3)$, as the standard tail (\ref{tail_non}) produced by the cubic
term. Combined together they give (see \cite{ym} for the derivation)
\begin{equation}
\phi(t,r) \approx \varepsilon^3 \phi_2(t,r)\sim \varepsilon^3\,c \,t^{-4}\,,\qquad
c=-8\int\limits_{-\infty}^{+\infty} a(u) {a'(u)}^2\,.
\end{equation}

Finally, we remark that although our analysis of tails was restricted to the flat background,
many conclusions carry over to more general asymptotically flat spacetimes, in particular black
hole spacetimes. For example, applying the ideas of section VII one can show that the massless
scalar field propagating outside a higher even-dimensional Schwarzschild black hole decays
anomalously fast as $\phi \sim t^{-(3d-5)}$ \cite{bcr}. This suggests that the problem of
asymptotic stability of the Schwarzschild black hole is easier in higher dimensions.

\subsection*{Acknowledgment:} I would like to thank Professor Andrzej Trautman for inviting me to
 give this
talk at the Banach Center.    This research was supported in part by the MNII grants 1PO3B01229
and SPB/189/6 PR EU/2007/7.

\end{document}